\documentclass{article}
\usepackage[utf8]{inputenc}
\usepackage{xcolor}
\usepackage{xspace}
\usepackage{fullpage}
\usepackage{times}
\usepackage{amsmath}
\usepackage{caption}

\DeclareCaptionType{equ}[Equation][]

\title{\color{DarkGreen}{Green AI}}
\author{Roy Schwartz\thanks{The first two authors contributed equally. The research was done at the Allen Institute for AI.}~~$^\diamondsuit$ \quad  Jesse Dodge$^{*\diamondsuit\clubsuit}$ \quad Noah A. Smith$^\diamondsuit$$^\heartsuit$ \quad Oren Etzioni$^\diamondsuit$\\
\\ $^\diamondsuit$Allen Institute for AI, Seattle, Washington, USA
\\ $^\clubsuit$ Carnegie Mellon University, Pittsburgh, Pennsylvania, USA
\\ $^\heartsuit$ University of Washington,  Seattle, Washington, USA
}
\date{July 2019}

\usepackage[numbers]{natbib}
\usepackage{graphicx}
\usepackage{booktabs}
\usepackage{tabularx}
\usepackage{hyperref}
\usepackage{wrapfig}
\usepackage{subfig}

\definecolor{DarkGreen}{RGB}{0,111,0}

\newcommand{\redai}{{\color{red}{Red AI}}\xspace}
\newcommand{\Redai}{{\color{red}{Red AI}}\xspace}
\newcommand{\greenai}{{\color{DarkGreen}{Green AI}}\xspace}
\newcommand{\Greenai}{{\color{DarkGreen}{Green AI}}\xspace}
\newcommand{\green}{{\color{DarkGreen}{green}}\xspace}

\newcommand{\red}{{\color{red}{red}}\xspace}

\newcommand{\com}[1]{}
\newcommand{\resolved}[1]{}

\newcommand{\figref}[1]{Figure~\ref{fig:#1}}
\newcommand{\secref}[1]{Section~\ref{sec:#1}}
\newcommand{\equref}[1]{Equation~\ref{eq:#1}}
\newcommand{\changeme}[1]{{\color{purple}{#1$_{\textrm{\it changeme}}$}}}
\newcommand{\add}{\textsc{add}\xspace}
\newcommand{\mul}{\textsc{mul}\xspace}
\newcommand{\flop}{{FPO}\xspace}

\newcommand{\eg}[0]{e.g.}
\newcommand{\Eg}[0]{E.g.}
\newcommand{\naipapers}{60\xspace}

\newcommand{\draftonly}[1]{#1}
\renewcommand{\draftonly}[1]{}
\newcommand{\draftcomment}[3]{\draftonly{\textcolor{#2}{[#3 ({\bf #1})]}}}
\newcommand{\roy}[1]{\draftcomment{Roy}{red}{#1}}
\newcommand{\nascomment}[1]{\draftcomment{Noah}{olive}{#1}}
\newcommand{\oren}[1]{\draftcomment{Oren}{blue}{#1}}
\newcommand{\jdcomment}[1]{\draftcomment{Jesse}{teal}{#1}}

\begin{document}

\maketitle

\begin{abstract}
    The computations required for deep learning research have been doubling every few months, resulting in an estimated 300,000x increase from 2012 to 2018 \cite{Amodei:2018}. These computations have a surprisingly large carbon footprint \cite{Strubell:2019}. Ironically, deep learning was inspired by the human brain, which is remarkably energy efficient.  Moreover, the financial cost of the computations can make it difficult for academics, students, and researchers, in particular those from emerging economies, to engage in deep learning research.
    
    This position paper advocates a practical solution by making {\bf efficiency} an evaluation criterion for research alongside accuracy and related measures. In addition, we propose reporting the financial cost or ``price tag'' of developing, training, and running models to provide baselines for the investigation of increasingly efficient methods.  Our goal is to make AI both greener and more inclusive---enabling any inspired undergraduate with a laptop to write high-quality research papers. \greenai is an emerging focus at the Allen Institute for AI.

\end{abstract}

\section{Introduction and Motivation}\label{sec:intro}

Since 2012, the  field  of artificial intelligence has reported remarkable progress on a broad range of capabilities including object recognition, game playing, machine translation, and more \cite{aiindex}.  This progress has been achieved  by increasingly large and computationally-intensive deep learning models.\footnote{For brevity, we refer to AI throughout this paper, but our focus is on AI research that relies on deep learning methods.}   \figref{openai} reproduced from \cite{Amodei:2018} plots \resolved{ \nascomment{accuracy?} against }training cost increase over time for state-of-the-art deep learning models starting with AlexNet in 2012 \cite{Krizhevsky:2012} to AlphaZero in 2017 \cite{Silver:2017a}.  The chart shows an overall increase of 300,000x, with training cost doubling every few months. An even sharper trend can be observed in NLP word embedding approaches by looking at ELMo \cite{Peters:2018} followed by BERT \cite{Devlin:2019}, openGPT-2 \cite{Radford:2019}, and XLNet \cite{Yang:2019}.  An important paper \cite{Strubell:2019} has estimated the carbon footprint of several NLP models and argued that this trend is both environmentally unfriendly (which we refer to as \redai) and  expensive,  raising  barriers to participation in NLP research.

\newcommand{\figlen}[0]{0.6\textwidth}

\begin{figure}[!t]
\center
\includegraphics[trim={0.5cm 1.75cm 0.5cm 2.75cm},clip,width=\figlen]{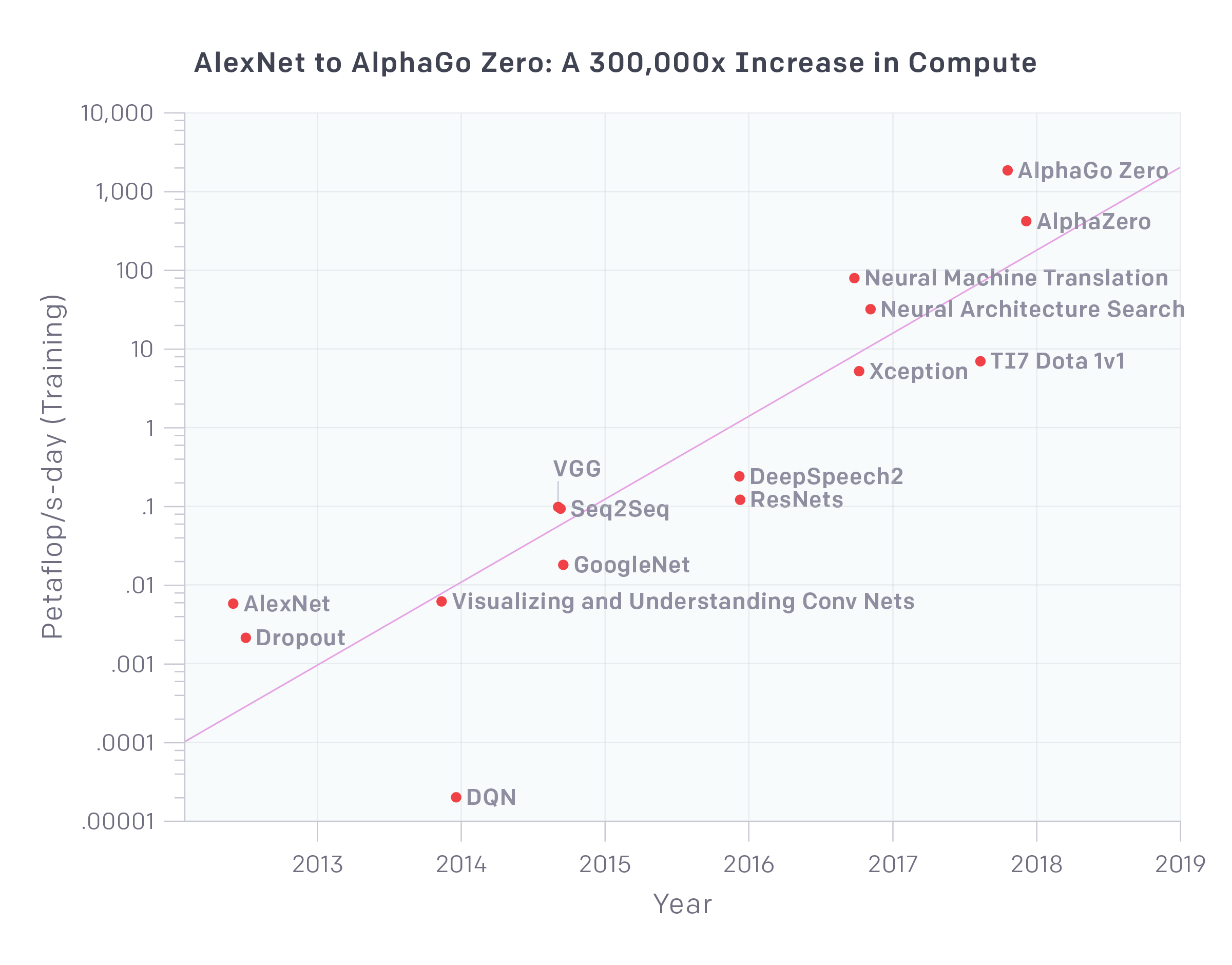}
\caption{\label{fig:openai} The amount of compute used to train deep learning models has increased 300,000x in 6 years. Figure taken from \cite{Amodei:2018}.}
\end{figure}

This trend is driven by the strong focus of the AI community on obtaining ``state-of-the-art'' results,\footnote{Meaning, in practice, that a system's accuracy on some benchmark is greater than any previously reported system's accuracy.}\resolved{\jdcomment{I added these hyphens, since I believe this is acting as a multi-word adjective, but please check}} as exemplified by the rising popularity of leaderboards \cite{Wang:2019a,Wang:2019b}, which typically report accuracy measures but omit any mention of cost or efficiency (see, for example, \href{https://leaderboards.allenai.org}{\ttfamily{leaderboards.allenai.org}}). 
Despite the clear benefits of improving model accuracy in AI, the focus on this single metric ignores the economic, environmental, or social cost of reaching the reported accuracy.

We advocate increasing research activity in \greenai---AI research that is more environmentally friendly and inclusive. 
We emphasize that \redai research has been yielding valuable contributions to the field of AI, but it's been overly dominant. 
We want to shift the balance towards the \greenai\ {\em option}---to ensure that any inspired undergraduate with a laptop has the opportunity to write high-quality papers that could be accepted at premier research conferences.
Specifically, we propose making {\bf efficiency} a more common evaluation criterion for AI papers alongside accuracy and related measures.

AI research can be computationally expensive in a number of ways, but each provides opportunities for efficient improvements; for example, papers could be required to plot accuracy as a function of computational cost and of training set size, providing a baseline for more data-efficient research in the future. 
Reporting the computational price tag of finding, training, and running models is a key \greenai\ practice (see \equref{greenai_equation}).  In addition to providing transparency, price tags are baselines that other researchers could improve on.

Our empirical analysis in \figref{recent_papers} suggests that the AI research community has paid relatively little attention to computational efficiency.  
In fact, as Figure \ref{fig:openai} illustrates, the computational cost of research is increasing exponentially, at a pace that far exceeds Moore's Law \cite{Moore:1965}. \redai\ is on the rise despite the well-known diminishing returns of increased cost (\eg, Figure \ref{fig:instagram_data}).  This paper identifies key factors that contribute to \redai\ and advocates the introduction of a simple, easy-to-compute efficiency metric that could help make some AI research greener, more inclusive, and perhaps more cognitively plausible.  \greenai\ is part of a broader, long-standing interest in environmentally-friendly scientific research (\eg, see the journal \emph{Green Chemistry}\resolved{\footnote{\url{https://www.rsc.org/journals-books-databases/about-journals/green-chemistry/}}}). Computer science, in particular, has a long history of investigating sustainable and energy-efficient computing (\eg, see the journal \emph{Sustainable Computing: Informatics and Systems}\resolved{\footnote{\url{https://www.journals.elsevier.com/sustainable-computing-informatics-and-systems}}}).  

The remainder of this paper is organized as follows. Section \ref{sec:redai} analyzes practices that move deep-learning research into the realm of \redai.  Section \ref{sec:greenai} discusses our proposals for \greenai.  Section \ref{sec:related} considers related work, and we conclude with a discussion of directions for future research.

\section{\Redai}\label{sec:redai}
\resolved{\emph{Science, and in particular  AI, advances in several different avenues. 
For instance, some of the biggest breakthroughs in AI resulted from the development of novel algorithms \cite{Hochreiter:1997,lecun_gradient-based_1998}, from observing analogies between AI components and phenomena from other domains \cite{Bahdanau:2015},
or from making novel connections between different AI components \cite{Peters:2018}.}
\roy{after writing the first paragraph I realize it might better fit the intro, so leaving it here in case it is useful elsewhere}
\nascomment{doesn't seem to fit currently}}

\redai\ refers to AI research that seeks to obtain state-of-the-art results in accuracy (or related measures) through the use of massive computational power---essentially ``buying" stronger results. 
Yet the relationship between model performance and model complexity (measured as number of parameters or inference time)
has long been understood to be at best logarithmic; for a linear gain in performance, an exponentially larger model is required \cite{Huang:2017}.
Similar trends exist with increasing the quantity of training data \cite{Sun:2017, Halevy:2009} and the number of experiments \cite{Dodge:2019}. In each of these cases, diminishing returns come at increased computational cost.

This section analyzes the factors contributing to \redai\ and shows how it is resulting in diminishing returns over time (see Figure \ref{fig:instagram_data}). 
We note again that \redai work is valuable, and in fact, much of it contributes to what we know by pushing the boundaries of AI. Our exposition here is meant to highlight areas where computational expense is high, and to present each as an opportunity for developing more efficient techniques.

\resolved{
While the the race for the state of the art on standard benchmarks has unequivocally contributed to the tremendous success of AI in recent years,
it has also created some troubling trends.
First, maximizing accuracy alone has encouraged rich organizations to put greater computational resources into developing models,  essentially ``buying'' stronger results.\resolved{\nascomment{I don't think the methods are red AI; the notion that accuracy is the only thing that matters, no matter the cost, is red AI.}}\resolved{\nascomment{not sure how to work it in, but I think we need to be even-handed in saying that there's nothing inherently wrong with leaderboards; they (or something like them) are a necessary ingredient in making progress across many institutions.  our point is that they are not \emph{sufficient}}}
Second, focusing on a single measure of success \com{
\redai has led to performance improvements, it }ignores the environmental toll of developing\resolved{ \nascomment{and tuning and executing?}} such models. 
Third, recent models have become so expensive to train, or even run, that they limit many researchers from working with them, thereby creating a high barrier for many people for doing cutting-edge research, as it is defined today.
As a result, we argue that maximizing accuracy alone is {\it insufficient},
and refer to this practice as \redai.}


To demonstrate the prevalence of \redai, we sampled \naipapers papers from top AI conferences (ACL,\footnote{\url{https://acl2018.org}} NeurIPS,\footnote{\url{https://nips.cc/Conferences/2018}} and CVPR\footnote{\url{http://cvpr2019.thecvf.com}}\com{, ICLR, EMNLP, NAACL?}\resolved{ \nascomment{ICML probably goes in that list; are there other equally high-profile vision conferences?}}).\com{\footnote{We randomly sampled \changeme{X} papers from the most recent event for each conference, and included only papers that had experimental results.}}
For each paper we noted whether the authors claim their main contribution to be (a) an improvement to accuracy or some related measure\resolved{\nascomment{too vague; this should more clearly defined as ``an improvement to accuracy or some related measure of model quality'' or something like that}}, (b) an improvement to efficiency\resolved{ \nascomment{no, not a measure, an improvement to efficiency}}, (c) both, or (d) other.\resolved{\jdcomment{i read the abstract and the conclusion, and looked at each figure and table in each paper. if they claimed even once that some variant of their approach was more efficient in some way than a baseline, i counted it as efficiency.}}
As shown in \figref{recent_papers}, in all conferences we considered, a large majority of the papers target accuracy (90\% of ACL papers, 80\% of NeurIPS papers and 75\% of CVPR papers).
Moreover, for both empirical AI conferences (ACL and CVPR) only a small portion (10\% and 20\% respectively) argue for a new efficiency result.\footnote{Interestingly, many NeurIPS papers included convergence rates or regret bounds which describe performance as a function of examples or iterations, thus targeting efficiency (55\%). This indicates an increased awareness of the importance of this concept, at least in theoretical analyses.}\resolved{\jdcomment{i would probably keep the part here about the the focus of most research being empirical results (e.g., through leaderboards, etc.) but i'd move the description of the Table towards the end.}}
This highlights the focus of the AI community on measures of performance such as accuracy, at the expense of measures of efficiency such as speed or model size. 
In this paper we argue that a larger weight should be given to the latter.

\resolved{We next turn to describe concrete suggestions on how to make AI more {\textcolor{DarkGreen}{green}}.}

\com{\begin{table}
\setlength{\tabcolsep}{4pt}
\centering
\begin{tabular}{l c c c c}
  \toprule
  {\bf Conference} & {\bf Accuracy} & {\bf Efficiency} & {\bf Both} & {\bf Other}   \\
  \midrule
  {\bf ACL 2018} & 16 & 0 & 2 & 2    \\
  {\bf CVPR 2019} & 13 & 2 & 2 & 3    \\
  {\bf NeurIPS 2018} & \textcolor{white}{0}9 & 4 & 7 & 1 \\

  \bottomrule
\end{tabular}}
\begin{figure}
\centering
\setlength{\belowcaptionskip}{-0.2cm}
\includegraphics[trim={0.1cm 0.1cm 0.1cm 0.1cm},clip,width=0.5\textwidth]{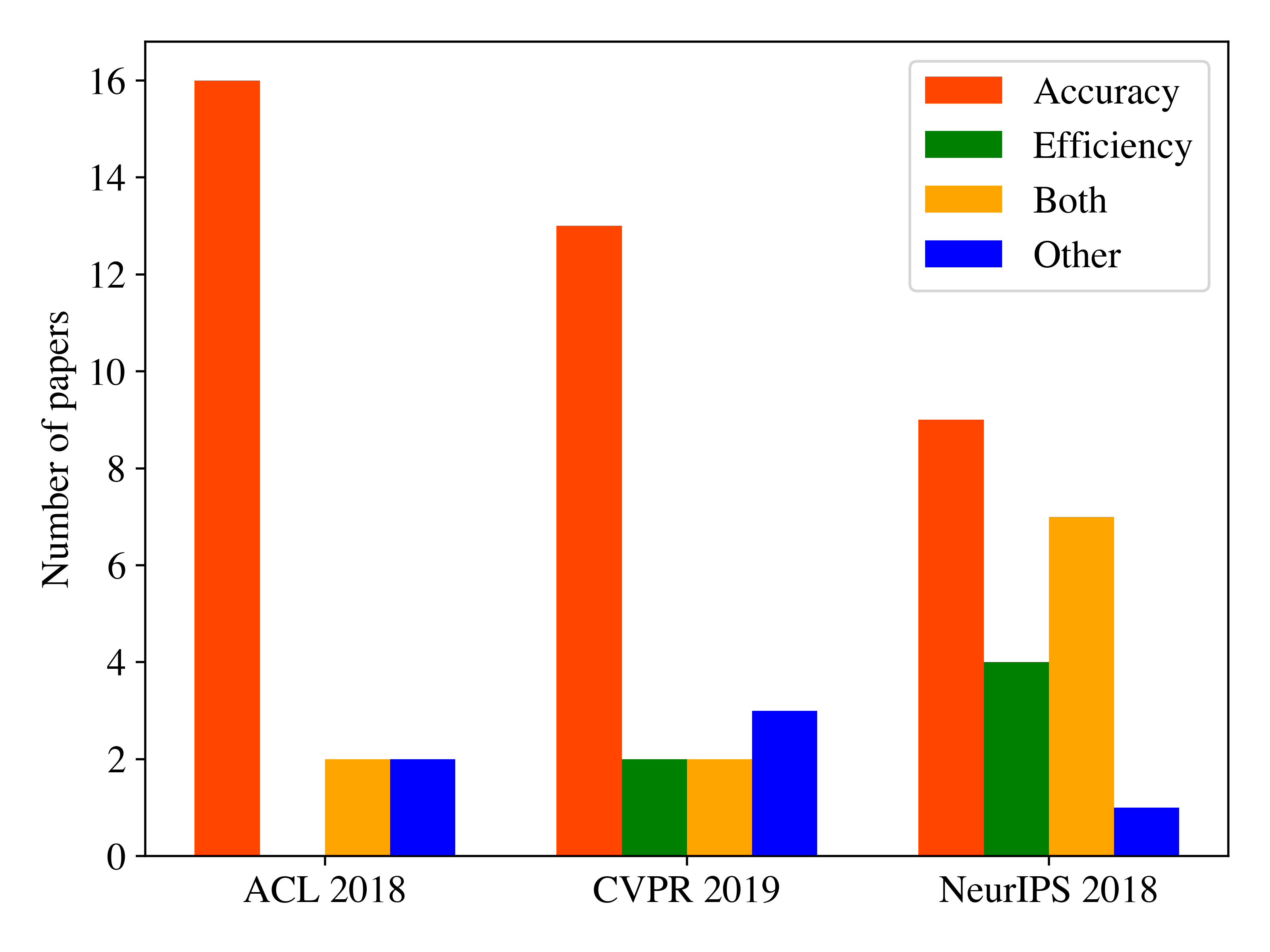}
\caption{\label{fig:recent_papers} AI papers tend to target \emph{accuracy} rather than \emph{efficiency}. 
The figure shows the proportion of papers that target accuracy, efficiency, both or other from a sample of \naipapers papers from top AI conferences.\resolved{\roy{How much effort should we put into this? is 4 conferences, $\sim$25 papers each sufficient?}}}
\end{figure}

To better understand the different ways in which AI research can be \red, consider an AI result reported in a scientific paper. 
This result typically includes a model trained on a training dataset and evaluated on a test dataset. 
The process of developing that model often involves multiple experiments to tune its hyperparameters. 
When considering the different factors that increase the computational and environmental cost of producing such a result, three factors come to mind: the cost of executing the model on a single ($E$)xample (either during training or at inference time); the size of the training ($D$)ataset, which controls the number of times the model is executed during training, and the number of ($H$)yperparameter experiments, which controls how many times the model is trained during model development. The total cost of producing a ($R$)esult in machine learning increases linearly with each of these quantities. This cost can be estimated as follows:

\begin{equ}[h]
\Large
\centering
\boxed{
{ Cost(R) \propto E \cdot D\cdot H}
}
\normalsize
\caption{\label{eq:greenai_equation}
The equation of \redai: The cost of an AI ($R$)esult grows linearly with the cost of processing a single ($E$)xample, the size of the training ($D$)ataset and the number of ($H$)yperparameter experiments.}
\end{equ}

\com{

The three most prominent factors are: using models for which it is expensive to process one example, applying models to massive datasets, and running numerous experiments. The total cost of getting a result in machine learning increases linearly with each of these quantities; if the cost of applying a model to a single example is $C$, and there are $N$ examples in a given dataset, and $K$ hyperparameter assignments (or model variants) are evaluated, the total cost can be estimated as follows:
}

\equref{greenai_equation} is a simplification (e.g., different hyperparameter assignments can lead to different costs for processing a single example). It also ignores other factors such as the number of training epochs. Nonetheless, it illustrates three quantities that are each an important factor in the total cost of generating a result.  Below, we  consider each quantity separately.

\resolved{, exemplifying the potential implications of each, while providing some prominent examples.}


\paragraph{Expensive processing of one example}
Our focus is on neural models, where it is common for each training step to require inference, so we discuss training and inference cost together as ``processing'' an example.
Some works have used increasingly expensive models which require great amounts of resources, and as a result, in these models, performing inference can require a lot of computation, and training even more so.\com{Models which have a massive number of parameters can be so large they don't fit on standard GPUs, requiring specialized hardware. 
Model size and processing cost both have diminishing returns for performance as we increase complexity; this relationship, combined with the push for improving state-of-the-art results, has led to some massively expensive models.}\resolved{to train and execute\resolved{ \nascomment{also to run?}}.\resolved{ \jdcomment{is it worth discussing models which are expensive because of inference, like some parsing models?}}}
For instance, Google's BERT-large \cite{Devlin:2019} contains roughly 350 million parameters.
openAI's openGPT2-XL model \cite{Radford:2019} contains 1.5 billion parameters. 
AI2, our home organization, recently released Grover \cite{Zellers:2019}, also containing 1.5 billion parameters.
In the computer vision community, a similar trend is observed (\figref{openai}).

Such large models have high costs for processing each example, which leads to large training costs. BERT-large was trained on 64 TPU chips for 4 days. Grover was trained on 256 TPU chips for two weeks, at an estimated cost of \$25,000. 
XLNet had a similar architecture to BERT-large, but used a more expensive objective function (in addition to an order of magnitude more data), and was trained on 512 TPU chips for 2.5 days.\footnote{Some estimates for the cost of this process reach \$250,000 (\href{https://twitter.com/eturner303/status/1143174828804857856}{\ttfamily{twitter.com/eturner303/status/1143174828804857856}}).}\com{most researchers, which cannot take part in contributing to state of the art (or even using these models).
Moreover, this process consumes large amounts of energy, and is thus unfriendly to the environment. }
\resolved{The size of these models not only limits their training but also their execution.}
It is impossible to reproduce the best BERT-large results\footnote{See \url{https://github.com/google-research/bert}} or XLNet results\footnote{See \url{https://github.com/zihangdai/xlnet}} using a single GPU. 
Specialized models can have even more extreme costs, such as AlphaGo, the best version of which required 1,920 CPUs and 280 GPUs to play a single game of Go \cite{Silver:2016} at a cost of over \$1,000 per hour.\footnote{Recent versions of AlphaGo are far more efficient\com{, e.g., AlphaGo Zero} \cite{Silver:2017b}.}

When examining variants of a single model (e.g., BERT-small and BERT-large) we see that larger models can have stronger performance, which is a valuable scientific contribution. However, this implies the financial and environmental cost of increasingly large AI models will not decrease soon, as the pace of model growth far exceeds the resulting increase in model performance \cite{Howard:2017}. As a result, more and more resources are going to be required to keep improving AI models by simply making them larger.

\paragraph{Processing many examples}
Another way state-of-the-art performance has recently been progressing in AI is by successively increasing the amount of training data models are trained on. 
BERT-large had top performance in 2018 across many NLP tasks after training on 3 billion word-pieces.
XLNet recently outperformed BERT after training on 32 billion word-pieces, including part of Common Crawl\com{at a similar size, openAI's }; openGPT-2-XL\com{ \cite{Radford:2019}} trained on 40 billion words; FAIR's RoBERTa \cite{Liu:2019} was trained on 160GB of text, roughly 40 billion word-pieces, requiring around 25,000 GPU hours to train.
In computer vision, researchers from Facebook \cite{Mahajan:2018} pretrained an image classification model on 3.5 billion images from Instagram, three orders of magnitude larger than existing labelled image datasets such as Open Images.\footnote{\url{https://opensource.google.com/projects/open-images-dataset}}\resolved{\roy{Check with vision folks: is this an extreme example? other key examples?}}

\com{ few works were trained models on massive amounts of data, which is internal and not shared with others. For instance, Google's original word2vec model \cite{Mikolov:2013} was trained on 100 Billion words from Google News, which isn't available outside Google.  Researchers from Facebook\resolved{  \nascomment{grammar problem?}} This dataset is not only very large in size, but is also unavailable  outside Facebook.}

\begin{figure}
\centering
\setlength{\belowcaptionskip}{-0.2cm}
\includegraphics[trim={0.1cm 0.1cm 0.1cm 0.1cm},clip,width=0.6\textwidth]{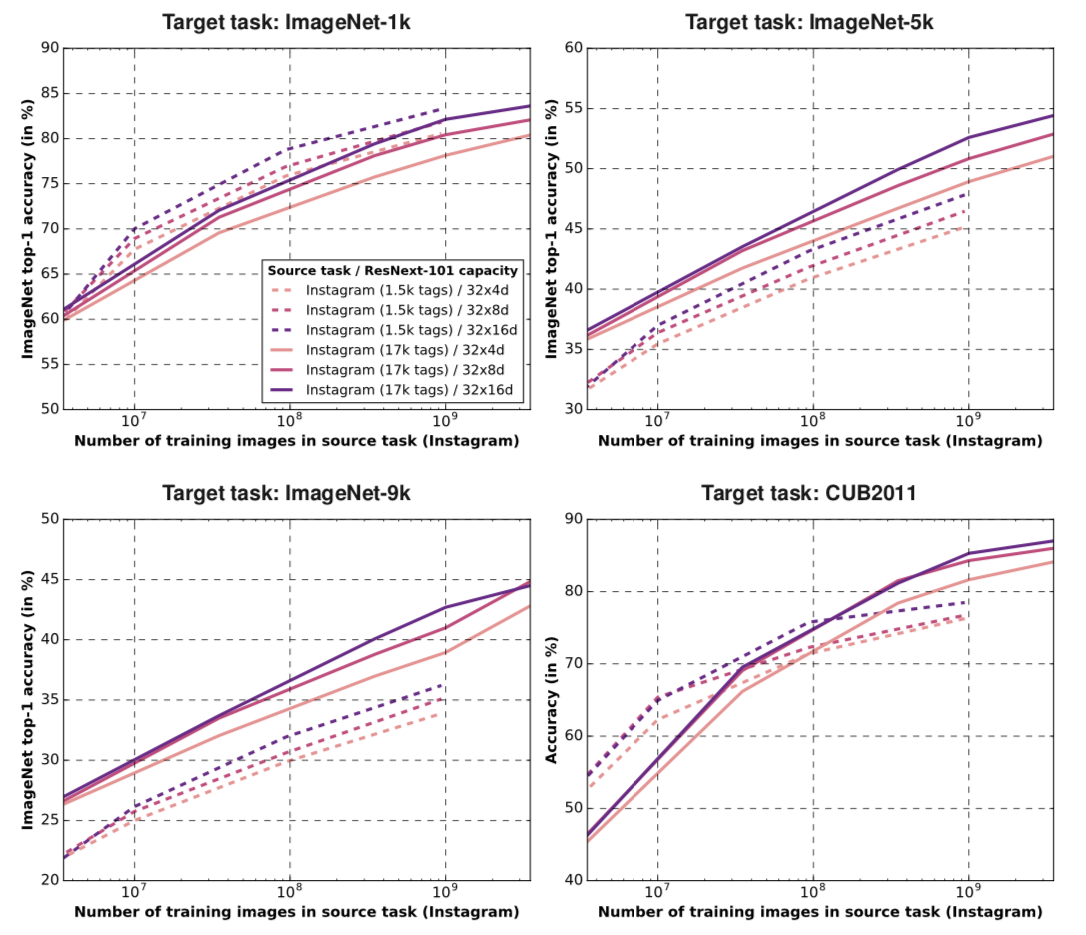}
\caption{\label{fig:instagram_data} Diminishing returns of training on more data: object detection accuracy increases linearly as the number of training examples increases exponentially \cite{Mahajan:2018}.\resolved{\roy{Oren: I changed the scale back to 90-100 because I thought it made the massage clearer: going from 10k to 20k yields a large jump. going from 40k to 80k yields a small jump. (also removed the exponential X scale as it doesn't help the visualization imho)}}
\resolved{\nascomment{should add a citation here}}\resolved{\roy{During a conversation with Rowan, he mentioned that the graph we used from his blog was not the right graph for our purposes, so I replaced it with this one}}}
\end{figure}

The use of massive data creates barriers for many researchers for reproducing the results of these models, or training their own models on the same setup (especially as training for multiple epochs is standard). For example, the June 2019 Common Crawl contains 242 TB of uncompressed data,\footnote{\url{http://commoncrawl.org/2019/07/}} so even storing the data is expensive.
Finally, as in the case of model size, 
relying on more data to improve performance is notoriously expensive because of the diminishing return of adding more data \cite{Sun:2017}. For instance, \figref{instagram_data}, taken from \cite{Mahajan:2018}, shows a logarithmic relation between the object recognition top-1 accuracy and the number of training examples.

\resolved{\nascomment{I think we should not emphasize the inaccessibility issue; that is another battle.  I think the more relevant points are (1) some of these datasets are so large that most groups could not replicate the methods \emph{even if they could get access to the data (or similar data)}, and (2) even using the models, when they are released, is too computationally expensive for many teams (I think this is true but you should tell me if I'm wrong) ... oh, sorry, that's your next point}}

\resolved{\nascomment{I feel like we're repeating the same points; if we said this in the intro, then here we need to back it up with more concrete numbers.  e.g., suppose I'm an academic.  I want to download BERT and fine-tune it on my task.  what will it cost me?  what kind of machine (or cloud instance) do I need to buy?}}

\resolved{\roy{@jesse missing: RL. For instance, Deepmind's AlphaGo was trained on 176 GPUs for 40 days \cite{Silver:2016}.}}

\paragraph{Massive number of experiments}\resolved{  \nascomment{I think ``search space'' is too ambiguous, maybe say ``Large design space for models''?}}
\resolved{\jdcomment{i would place this third, it's perhaps the least intuitive for a casual reader. i think either 'large data' or 'large model' could be first.}}
Some projects have poured large amounts of computation into tuning hyperparameters or searching over neural architectures, well beyond the reach of most researchers.
For instance, \resolved{\nascomment{\resolved{again, weird grammar -- }is it important to attribute this to Google?}\roy{good question! I think it highlights that the strong/rich groups are responsible for red AI. But it is a bit aggressive to constantly blame google et al.. also note that many of these papers have academic collaborators which we do not mention. thoughts on this?}}researchers from Google \cite{Zoph:2017} trained over 12,800 neural networks in their neural architecture search to improve performance on object detection and language modeling.
With a fixed architecture, researchers from DeepMind \cite{Melis:2018} evaluated 1,500 hyperparameter assignments to demonstrate that an LSTM language model \cite{Hochreiter:1997} can reach state-of-the-art perplexity results.\nascomment{the big numbers don't really tell the whole story, can we make this feel more real by putting it into FLOPs and dollars?  as before}\roy{jesse?}\jdcomment{i'm working on estimating cost or FLOPs here correct.}
Despite the value of this result in showing that the performance of an LSTM does not plateau after only a few hyperparameter trials, fully exploring the potential of other competitive models for a fair comparison is prohibitively expensive. 

The topic of massive number of experiments is not as well studied as the first two discussed above. In fact, the number of experiments performed during model construction is often underreported. Nonetheless, evidence  for a logarithmic relation exists here as well, between the number of experiments and performance gains \cite{Dodge:2019}. 

\com{
While the amount of computation available has increased gradually in the past century \cite{Moore:1965}, the inequality of the distribution of resources in the AI community has seen a surge of results from strong and rich research labs that gradually improve the state of the art by putting more resources into model development, be it time, memory or data.
While some of these improvements have undoubtedly moved the field forward, 
}

\paragraph{Discussion}
The benefits of pouring more resources into models are certainly of interest to the AI community.
Indeed, there is value in pushing the limits of model size, dataset size, and the hyperparameter search space. Currently, despite the massive amount of resources put into recent AI models, such investment still pays off in terms of downstream performance (albeit at an increasingly lower rate). Finding the point of saturation (if such exists) is an important question for the future of AI. 

Our goal in this paper is to raise awareness of the cost of \redai, as well as encourage the AI community to recognize the value of work by researchers that take a different path, optimizing efficiency rather than accuracy.
Next we turn to discuss concrete measures for making AI more \green.

\section{\Greenai}\label{sec:greenai}

The term \greenai\ refers to AI research that yields novel results without increasing computational cost, and ideally reducing it.  Whereas \redai\ has resulted in rapidly escalating computational (and thus carbon) costs, \greenai\ has the opposite effect. If measures of efficiency are widely accepted as important evaluation metrics for research alongside accuracy, then researchers will have the option of focusing on the efficiency of their models with positive impact on both the environment and inclusiveness.
This section reviews several measures of efficiency that could be reported and optimized, and advocates one particular measure---\flop---which we argue should be reported when AI research findings are published.  

\subsection{Measures of Efficiency}\label{sec:measures}

\resolved{\nascomment{Maybe list our desiderata up front?  E.g., we want to count something that is stable across different labs, different times, etc.}  another point still missing is the scope of the measurement:  one round of inference?  training one model?  hyperparameter tuning?  the whole paper?}
\resolved{\roy{implementation dependence}}

To measure efficiency, we suggest reporting the amount of work required to generate a result in AI, that is, the amount of work required to train a model, and if applicable, the sum of works for all hyperparameter tuning experiments.
As the cost of an experiment decomposes into the cost of a processing a single example, the size of the dataset, and the number of experiments (\equref{greenai_equation}), reducing the amount of work in each of these steps will result in AI that is more \green.\resolved{\roy{note the change here and after. Given the green AI equation, I changed \fpo to measure the work done in order to get a result, not per inference.}}

When reporting the amount of work done by a model, we want to measure a quantity that allows for a fair comparison between different models. As a result, this measure should ideally be stable across different labs, at different times, and using different hardware.

\paragraph{Carbon emission}
Carbon emission is appealing as it is a quantity we want to directly minimize. Nonetheless it is impractical to measure the exact amount of carbon released by training or executing a model, and accordingly---generating an AI result, as this amount depends highly on the local electricity infrastructure. As a result, it is not comparable between researchers in different locations or even the same location at different times.

\paragraph{Electricity usage}
Electricity usage is correlated with carbon emission while being time- and location-agnostic. 
Moreover, GPUs often report the amount of electricity each of their cores consume at each time point, which facilitates the estimation of the total amount of electricity consumed by generating an AI result.
Nonetheless, this measure is hardware dependent, and as a result does not allow for a fair comparison between different models\resolved{\nascomment{for some other reason or just because of hardware dependence?}}.

\paragraph{Elapsed real time}
The total running time for generating an AI result is a natural measure for efficiency, as all other things being equal, a faster model is doing less computational work. 
Nonetheless, this measure is highly influenced by factors such as the underlying hardware, other jobs running on the same machine, and the number of cores used. 
These factors hinder the comparison between different models, as well as the decoupling of modeling contributions from hardware improvements.\resolved{\nascomment{do we want to talk, somewhere, about implementation differences?  this is arguably a bit of a weakness for FLOP, too, but I think it's better to discuss than avoid}}

\paragraph{Number of parameters}
Another common measure of efficiency is the number of parameters (learnable or total) used by the model. 
As with run time, this measure is correlated with the amount of work.
Unlike the other measures described above, it does not depend on the underlying hardware.
Moreover, this measure also highly correlates with the amount of memory consumed by the model. 
Nonetheless, different algorithms make different use of their parameters, for instance by making the model deeper vs.~wider. 
As a result, different models with a similar number of parameters often perform different amounts of work.

\paragraph{\flop}
As a concrete measure, we suggest reporting the total number of floating point operations (\flop) required to generate a result.\footnote{Floating point operations are often referred to as FLOP(s), though this term is not uniquely defined \cite{Gordon:2018}.\resolved{\roy{Jesse, examples of different definitions?}.} To avoid confusion, we use the term \flop.}\resolved{\jdcomment{can we call it FLOP cost?}}
\flop provides an estimate to the amount of work performed by a computational process.\resolved{\nascomment{not really; it's a quantity, not a way to estimate it!}}
It is computed analytically by defining a cost to two base operations, \add and \mul. 
Based on these operations, the \flop cost of any machine learning\resolved{ \nascomment{not just DL}} abstract operation (e.g., a tanh operation, a matrix multiplication, a convolution operation, or the BERT model)  can be computed as a recursive function of these two operations. 
\flop has been used in the past to quantify the energy footprint of a model \cite{Molchanov:2017,Veniat:2018,Gordon:2018,Vaswani:2017}, but is not widely adopted in\resolved{ many \nascomment{any?} sub-fields of} AI. 

\resolved{\nascomment{I think we can clean this up by separating ``FLOP''-- the thing we think people should count---from the abstract thing we want measured (computational work), and also from the method for estimating FLOP cost.}}

\flop has several appealing properties. First, it directly computes the amount of work done by the running machine when executing a specific instance of a model, and is thus tied to the amount of energy consumed. 
Second, \flop is agnostic to the hardware on which the model is run\resolved{\jdcomment{i'd put this first, though it's not bad in this order}}. This facilitates fair comparisons between different approaches, unlike the measures described above.
Third, \flop is strongly correlated with the running time of the model \cite{Canziani:2017}. Unlike asymptotic runtime, \flop also considers the amount of work done at each time step. \resolved{\nascomment{should also talk explicitly about the pitfalls of reporting wall time.  should also talk about the weaknesses of what we're proposing.  e.g., in the past we used to shy away from reporting measures of model runtime because we didn't always bother to optimize our code; a lot depended on implementation.  I think we need to talk about that issue head-on and say why we think it's time to reverse that norm.}}

Several packages exist for computing FPO in various neural network libraries,\footnote{\Eg, \url{https://github.com/Swall0w/torchstat} ; \url{https://github.com/Lyken17/pytorch-OpCounter}} though none of them contains all the building blocks required to construct all modern AI models. We encourage the builders of neural network libraries to implement such functionality directly.
\resolved{\roy{@jesse: missing: existing packages, their limitations}\jdcomment{is a footnote the right way to cite these? i can add to references if that's better.}}

\resolved{\nascomment{*** might make sense to rework this section in a way that lays out all the options we know about for measuring/reporting what we want to measure/report, with pros and cons.  then explain why we advocate for measuring floating point operations per whatever, in today's world.}}

\paragraph{Discussion}

Efficient machine learning approaches have received attention in the research community, but are generally not motivated by being \green.
For example, a significant amount of work in the computer vision community has addressed efficient inference, which is necessary for real-time processing of images for applications like self-driving cars \cite{Ma:2018, Rastegari:2016, Liu:2016}, or for placing models on devices such as mobile phones \cite{Howard:2017, Sandler:2018}.
Most of these approaches target efficient model inference \cite{Redmon:2016,Zhang:2018,Gordon:2018},\footnote{Some very recent work also targeted efficient training \cite{Dettmers:2019}.} and thus only minimize the cost of processing a single example, while ignoring the other two \red practices discussed in \secref{redai}.\footnote{In fact, creating smaller models often results in longer running time, so mitigating the different trends might be at odds \cite{Walsman:2019}.}

\resolved{
\oren{I moved this sentence here from the intro to improve flow and reduce repetition--we're saying the same thing as above---please unify}
Research in efficient computer vision models has recently gained increased attention, focusing on small devices such as smartphones or real-time applications such as self-driving cars \cite[\emph{inter alia}]{Redmon:2016,Rastegari:2016,Zhang:2018,Gordon:2018}, but with less attention to the environmental aspect.
}


The above examples indicate that the path to making AI \green depends on how it is used. When developing a new model, much of the research process involves training many model variants on a training set and performing inference on a small development set. 
In such a setting, more efficient training procedures can lead to greater savings, while in a production setting more efficient inference can be more important.
We advocate for a holistic view of computational savings which doesn't sacrifice in some areas to make advances in others.

\flop has some limitations. First, it targets the electricity consumption of a model, while ignoring other potential limiting factors for researchers such as the memory consumption by the model, which can often lead to additional energy and monetary costs \cite{Ma:2018}.
Second, \com{as noted earlier, \flop also depends on the specific implementation, which is not typically considered part of the scientific contribution. We believe it is time it became so.\roy{too bold? anything else we want to acknowledge?}
\resolved{\nascomment{what about memory?  we're ignoring that; should say something}}

W}the amount of work done by a model largely depends on the \emph{model implementation}, as two different implementations of the same model could result in very different amounts of processing work. Due to the focus on the modeling contribution, the AI community has traditionally ignored the quality or efficiency of models' implementation.\footnote{We consider this exclusive focus on the final prediction  another symptom of \redai.} We argue that the time to reverse this norm has come, and that exceptionally good implementations that lead to efficient models should be credited by the AI community.

\resolved{Minimizing the amount of work for either case (training, inference or hyper-parameter search) will result in AI that is more \green.\resolved{\nascomment{We suggest reporting the FLOP cost for training a model, performing inference, of inference, training, and hyperparameter tuning?  I think if we're not careful we're letting people off too easy.  All three of these matter, and provide possible research directions for improvements.}}}
\resolved{\roy{Jonathan Bisk's paper}}

\com{The amount of computational work of a model highly depends on the specific implementation; an optimized implementation can result in a much more efficient model.
In the past, AI research didn't value model optimization, as the only thing that mattered was the final score. We consider this to be another symptom of \redai, and believe it's time to reverse that norm.}
\resolved{\nascomment{should also talk about the weaknesses of what we're proposing.  e.g., in the past we used to shy away from reporting measures of model runtime because we didn't always bother to optimize our code; a lot depended on implementation.  I think we need to talk about that issue head-on and say why we think it's time to reverse that norm.}}

\subsection{\flop Cost of Existing Models}

\begin{figure}
\centering
\renewcommand{\figlen}[0]{0.49\textwidth}
\subfloat[Different models.\label{fig:vision_flop_a}]{
\com{
\begin{tabular}{l c c c c }
  \toprule
  {\bf Name}& {\bf params (M)} & {\bf \flop (B)}& {\bf acc.}& {\bf year}     \\
  \midrule
AlexNet& \phantom{0}61.1& \phantom{0}0.7& 56.4& 2012    \\
ResNet152& \phantom{0}60.2& 11.6& 78.4& 2015    \\
ResNext& \phantom{0}83.5& 15.5& 79.0& 2017    \\
DPN107& \phantom{0}86.9& 18.4& 79.7& 2017    \\
SENet154& 115.1& 20.8& 81.3& 2018    \\
  \bottomrule
\end{tabular}
}
\includegraphics[trim={.8cm 0cm 1cm .5cm},clip,width=\figlen]{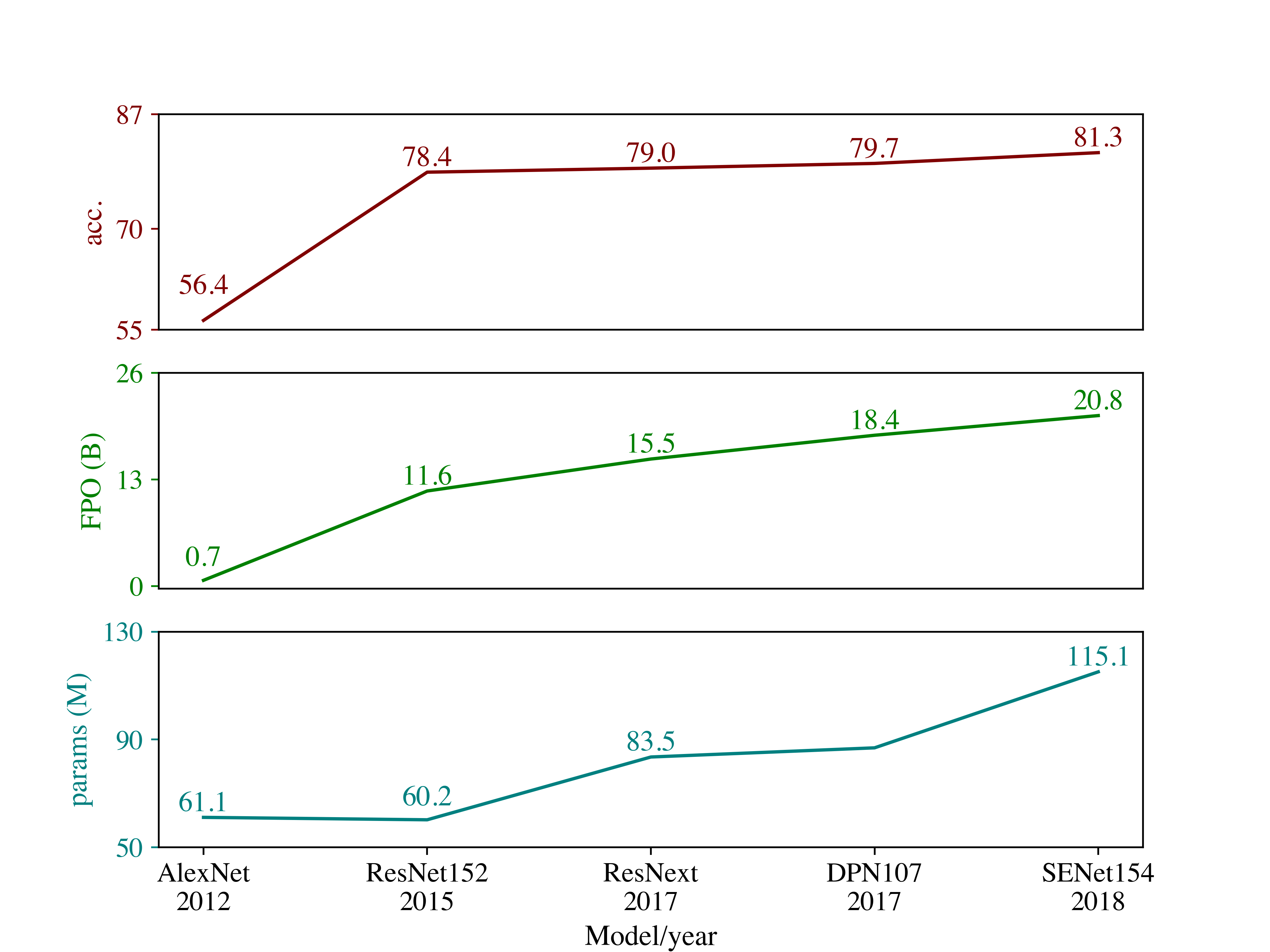}
}
\subfloat[Different layers of the ResNet model.\label{fig:vision_flop_b}]{
\com{\begin{tabular}{l c c c}
  \toprule
  {\bf layers}& {\bf params (M)} & {\bf \flop (B)}& {\bf acc.} \\
  \midrule
18	& 11.7 & \phantom{0}1.8	& 70.1	\\
34	& 21.8& 	\phantom{0}3.7& 	73.6\\
50	& 25.6& 	\phantom{0}4.1&	76.0\\
  101	& 44.5& 	\phantom{0}7.8&	77.4\\
152	& 60.2	& 11.6&	78.4\\
  \bottomrule
\end{tabular}
}
\includegraphics[trim={.8cm 0cm 1cm .5cm},clip,width=\figlen]{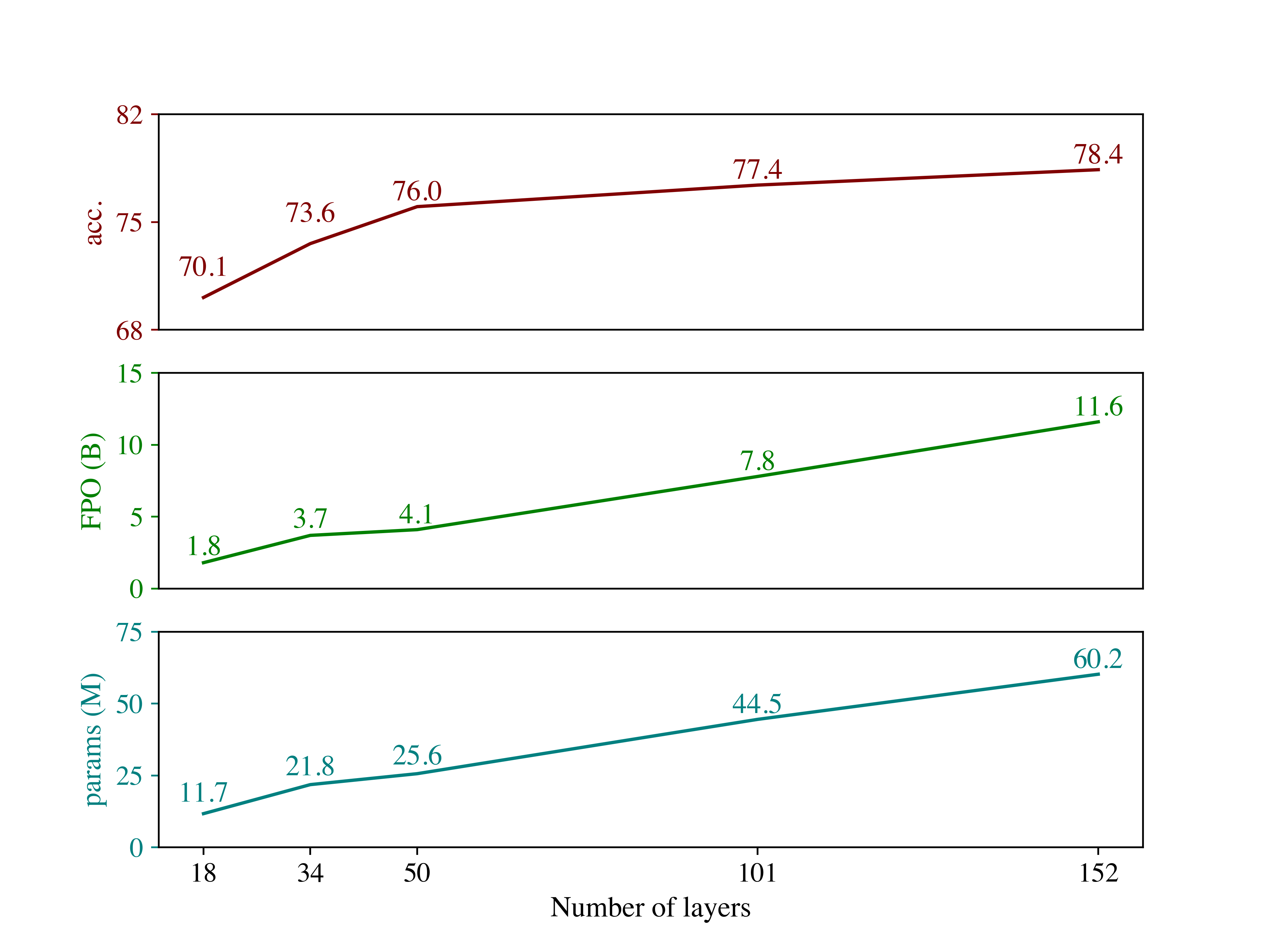}
}
\caption{\label{tab:vision_flop}
\jdcomment{this is missing some rows, so it's a little preliminary to really analyze.} Increase in \flop results in diminishing return for object detection top-1 accuracy. 
Plots (bottom to top): model parameters (in million), \flop (in billions), top-1 accuracy on ImageNet.\resolved{\nascomment{just move cites into table; don't need subcaption here}} (\ref{fig:vision_flop_a}): Different models: AlexNet \cite{Krizhevsky:2012}, ResNet \cite{He:2016}, ResNext \cite{Xie:2017}, DPN107 \cite{Chen:2017}, SENet154 \cite{Hu:2018}.
(\ref{fig:vision_flop_b}): Comparison of different sizes (measured by the number of layers) of the ResNet model \cite{He:2016}. 
}
\end{figure}

\resolved{\nascomment{in table headers, I think ``\#\fpo'' is redundant, why not just ``\fpo''?}}

To demonstrate the importance of reporting the amount of work, we present \flop costs for several existing models.\footnote{These numbers represent \flop per inference, i.e., the work required to process a single example.}\resolved{ \nascomment{the more I read this, the weirder it sounds.  I think we need a different name for measuring/reporting computational cost that is not the same as the unit we report it in.}}
\figref{vision_flop_a} shows the number of parameters and \flop of several leading object recognition models, as well as their performance on the ImageNet dataset \cite{Deng:2009}.\footnote{Numbers taken from \url{https://github.com/sovrasov/flops-counter.pytorch}}
\resolved{While these models have different architectures and are thus not directly comparable \nascomment{wait, why not?  we compare different architectures all the time!}, a}A few trends are observable.
First, as discussed in \secref{redai}, models get more expensive with time, but the increase in \flop does not lead to similar performance gains. For instance, an increase of almost 35\% in \flop between ResNet and ResNext (second and third points in graph) resulted in a 0.5\% top-1 accuracy improvement. Similar patterns are observed when considering the effect of other increases in model work.
Second, the number of model parameters does not tell the whole story: AlexNet (first point in the graph) actually has more parameters than ResNet (second point), but dramatically less \flop, and also much lower accuracy.\resolved{\nascomment{here and elsewhere, we need to talk about these models and their performance in a way that doesn't assume the reader knows all the acronyms and benchmarks!}}

\figref{vision_flop_b} shows the same analysis for a single object recognition model, ResNet \cite{He:2016}, while comparing different versions of the model with different number of layers. This creates a controlled comparison between the different models, as they are identical in architecture, except for their size (and accordingly, their \flop cost).
Once again, we notice the same trend: the large increase in \flop cost does not translate to a large increase in performance.\roy{Missing: NLP example}

\resolved{\nascomment{maybe better to present these in a plot; diminishing returns will be more visible that way.}}

\com{
\begin{table}
\setlength{\tabcolsep}{5.2pt}

\centering
\begin{tabular}{l c c c}
  \toprule
  {\bf \#layers}& {\bf \#params (M)} & {\bf \#\flop (B)}& {\bf acc.} \\
  \midrule
18	& 11.7 & \phantom{0}1.8	& 70.1	\\
34	& 21.8& 	\phantom{0}3.7& 	73.6\\
50	& 25.6& 	\phantom{0}4.1&	76.0\\
  101	& 44.5& 	\phantom{0}7.8&	77.4\\
152	& 60.2	& 11.6&	78.4\\
  \bottomrule
\end{tabular}
\caption{\label{tab:resnet_flop}
Comparison of different sizes (measured by the number of layers) of the ResNet model. top-1 accuracy is measured on ImageNet.}
\end{table}
}

\subsection{Additional Ways to Promote \greenai}


In addition to reporting the \flop cost of the final reported number, we encourage researchers to report the budget/accuracy curve observed during training. 
In a recent paper \cite{Dodge:2019}, we observed that selecting the better performing model on a given task depends highly on the amount of compute available during model development. We introduced a method for computing the expected best validation performance of a model as a function of the given budget. We argue that reporting this curve will allow users to make wiser decisions about their selection of models and highlight the stability of different approaches. 

We further advocate for making efficiency an official contribution in major AI conferences, by advising reviewers to recognize and value contributions that do not strictly improve state of the art, but have other benefits such as efficiency. 
Finally, we note that the trend of releasing pretrained models publicly\com{ (e.g., Google's BERT and XLNet)} is a \green success, and we would like to encourage organizations to continue to release their models in order to save others the costs of retraining them\com{, and in particular commend companies and organization who release trained models when it can be logistically difficult}.
\resolved{\nascomment{still murky:  can anyone really use these big models?}\roy{We discussed models that cannot be run earlier and will discuss them in the next paragraph, but I don't think we would be better off if Google didn't release XLNet because it's too big to fit on a single GPU}}

\section{Related Work}\label{sec:related}

Recent work has analyzed the carbon emissions of training deep NLP models \cite{Strubell:2019} and concluded that computationally expensive experiments can have a large environmental and economic impact. 
With modern experiments using such large budgets, many researchers (especially those in academia) lack the resources to work in many high-profile areas; increased value placed on computationally efficient approaches will allow research contributions from more diverse groups. 
\resolved{\nascomment{reworded here for clarity} }We emphasize that the conclusions of \cite{Strubell:2019} are the result of long-term trends, and are not isolated within NLP, but hold true across machine learning.

While some companies offset electricity usage by purchasing carbon credits, it is not clear that buying credits is as effective as using less energy. In addition, purchasing carbon credits is voluntary; Google cloud\footnote{\url{https://cloud.google.com/sustainability/}} and Microsoft Azure\footnote{\url{https://www.microsoft.com/en-us/environment/carbon}} purchase carbon credits to offset their spent energy, but Amazon's AWS\footnote{\url{https://aws.amazon.com/about-aws/sustainability/}} (the largest cloud computing platform\footnote{\url{https://tinyurl.com/y2kob969}}) only covered fifty percent of its power usage with renewable energy. 

The push to improve state-of-the-art performance has focused the research community's attention on reporting the single best result after running many experiments for model development and hyperparameter tuning. 
Failure to fully report these experiments prevents future researchers from understanding how much effort is required to reproduce a result or extend it \cite{Dodge:2019}.

Our focus is on improving efficiency in the machine learning community, but machine learning can also be used as a tool for work in areas like climate change.
For example, machine learning has been used for reducing emissions of cement plants \cite{Acharyya:2019} and tracking animal conservation outcomes \cite{Duhart:2019}, and is predicted to be useful for forest fire management \cite{Rolnick:2019}.
Undoubtedly these are important applications of machine learning; we recognize that they are orthogonal to the content of this paper.

\section{Conclusion}

The vision of \greenai\ raises many exciting research directions that help to overcome the inclusiveness challenges of \redai.
Progress will reduce the computational expense with a minimal reduction in performance, or even improve performance as more efficient methods are discovered.
Also, it would seem that \greenai\ could be moving us in a more cognitively plausible direction as the brain is highly efficient.

It's important to reiterate that we see \greenai\ as a valuable {\em option} not an exclusive mandate---of course, both \greenai and \redai have
contributions to make.  We want to increase the prevalence of \greenai\ by highlighting its benefits, advocating a standard measure of efficiency.
Below, we point to a few important \green research directions, and highlight a few open questions.

Research on building {\bf space} or {\bf time} efficient models is often motivated by fitting  a model on a small device (such as a phone) or fast enough to process examples in real time, such as image captioning for the blind (see \secref{measures})\resolved{\roy{Add vision works to related work}}. Some modern models don't even fit on a single GPU (see \secref{redai}).  Here we argue for a far broader approach.

{\bf Data efficiency} has received significant attention over the years \cite{Schwartz:2018,Kamthe:2018}.\roy{earlier examples?@jesse?}
Modern research in vision and NLP often involves first pretraining a model on large ``raw'' (unannotated) data then fine-tuning it to a task of interest through supervised learning.
A strong result in this area often involves achieving similar performance to a baseline with fewer training examples or fewer gradient steps. 
Most recent work has addressed fine-tuning data \cite{Peters:2018}\resolved{\nascomment{cites}}, but pretraining efficiency is also important.
In either case, one simple technique to improve in this area is to simply report performance with different amounts of training data.
For example, reporting performance of contextual embedding models trained on 10 million, 100 million, 1 billion, and 10 billion tokens would facilitate faster development of new models, as they can first be compared at the smallest data sizes.
Research here is of value not just to make training less expensive, but because in areas such as low resource languages or historical domains it is extremely hard to generate more data, so to progress we must make more efficient use of what is available.\com{For example, \cite{Yang:2019} pretrained on Common Crawl (after filtering it ``aggressively''), which is a copy of the Internet, so it's hard to get more data. \nascomment{devil's advocate:  the internet is always growing ...  not sure we want to get into a debate about the ceiling of big data}}\resolved{ \nascomment{another situation where you want to pretrain on smaller amounts of data has to do with domain shift, or wanting the ability to curate the data your method is exposed to, even in pretraining}}

Finally, the total number of experiments run to get a final result is often underreported and underdiscussed \cite{Dodge:2019}. The few instances researchers have of full reporting of the hyperparameter search, architecture evaluations, and ablations that went into a reported experimental result have surprised the community \cite{Strubell:2019}. While many  hyperparameter optimization algorithms exist which can reduce the computational expense required to reach a given level of performance \cite{Bergstra:2011,Dodge:2017}, simple improvements here can have a large impact. For example, stopping training early for models which are clearly underperforming can lead to great savings \cite{li:2017}.

\bibliographystyle{plain}
\bibliography{references}
\end{document}